\documentclass[acmsmall,screen]{acmart}
\usepackage{graphicx}
\usepackage{hyperref}
\usepackage{amsmath}
\usepackage{enumitem}
\usepackage{array}
\usepackage{svg}
\usepackage{wrapfig}
\usepackage{multirow}
\usepackage[colorinlistoftodos]{todonotes}
\usepackage{subcaption}
\usepackage[bb=boondox,bbscaled=.95,cal=boondoxo]{mathalfa}
\usepackage{array}
\newcolumntype{P}[1]{>{\raggedright\arraybackslash}p{#1}}

\newcommand{\Cli}{\textsc{Client}}
\newcommand{\Spa}{\textsc{Space}}
\newcommand{\Ser}{\textsc{Server}}
\newcommand{\Agg}{\textsc{Aggregator}}
\newcommand{\Act}{\textsc{Action}}
\newcommand{\Fee}{\textsc{Feed}}
\newcommand{\SSt}{\textsc{Space Storage}}
\newcommand{\ISt}{\textsc{Identity Storage}}
\newcommand{\SCu}{\textsc{Space Curator}}
\newcommand{\PCu}{\textsc{Personal Curator}}

\AtBeginDocument{
  }

\setcopyright{rightsretained}
\copyrightyear{2026}
\acmYear{2026}
\acmDOI{XXXXXXX.XXXXXXX}
\acmConference[CSCW 'XX]{Conference TItle}{Month 03--05,
  2026}{Location TBD}
\acmISBN{978-1-4503-XXXX-X/XXXX/XX}

\acmSubmissionID{4033}

\begin{document}

\title{Seeing the Politics of Decentralized Social Media Protocols}

\author{Tolulope Oshinowo}
\affiliation{
  \institution{Princeton University}
  \city{Princeton}
  \state{New Jersey}
  \country{USA}
}

\author{Sohyeon Hwang}
\affiliation{
  \institution{Princeton University}
  \city{Princeton}
  \state{New Jersey}
  \country{USA}
}

\author{Amy X. Zhang}
\affiliation{
  \institution{University of Washington}
  \city{Seattle}
  \state{Washington}
  \country{USA}
}

\author{Andrés Monroy-Hernández}
\affiliation{
  \institution{Princeton University}
  \city{Princeton}
  \state{New Jersey}
  \country{USA}
}

\begin{abstract}
Calls to decentralize feed-based social media have been driven by concerns about the concentrated power of centralized platforms and their societal impact. In response, numerous decentralized social media protocols have emerged, each interpreting ``decentralization'' in different ways. We analyze four such protocols---ActivityPub, AT Protocol, Nostr, and Farcaster---to develop a novel conceptual framework for understanding how protocols operationalize decentralization. Drawing from protocol documentation, media coverage, and first-hand interviews with protocol developers and experts, we contextualize each protocol's approach within their respective socio-technical goals. Our framework highlights how control over key components is distributed differently across each protocol, shaping who holds power over what kinds of decisions. How components are arranged in relation to one another further impacts how component owners might offset each other's power in shaping social media. We argue that examining protocols as artifacts reveals how values shape infrastructure and power dynamics---and that with a holistic framework as a guide, we can more effectively evaluate and design decentralized platforms aligned with the social and political futures we envision.
\end{abstract}




\maketitle
\section{Introduction}
The landscape of social media has been dominated by centralized platforms that control the flow of information, user data, and governance structures \cite{gillespie_custodiansinternet_2018, massanari_gamergatefappening_2017, crawford_whatflag_2016}. These platforms---such as Facebook, Instagram, and $\mathbb{X}$ (formerly Twitter)---have collectively shaped digital experiences for billions of people, but have also raised concerns about privacy, power imbalances, and algorithmic biases \cite{mauran_metausing_2024, brice_howstop_2024, lustig_algorithmicauthority_2016, duffy_platformgovernance_2023, diasoliva_fightinghate_2021}. In response, new decentralized social media protocols have surged in popularity \citep{cava_driverssocial_2023, jeong_exploringplatform_2024, nunes_usersentiments_2023}. These protocols aim to disseminate power to users, reduce reliance on centralized entities, and empower a rich ecosystem of interoperable social media services \cite{karppi_ifnot_2023, gehl_casealternative_2015, masnick_protocolsnot_2019}.

As decentralized social media protocols gain traction, researchers have turned their attention to understanding the user experience in these emerging ecosystems: tracking user behavior following moments of mass migration \citep{jeong_exploringplatform_2024, he_flockingmastodon_2023, cava_driverssocial_2023, nunes_usersentiments_2023} and identifying potential privacy or safety concerns \citep{laux_trustelephant_2023,cramer_trustsafety_2024}. Underpinning this work is whether decentralized social media protocols can fulfill their promise of user agency and control \citep{dunbar-hester_showingyour_2024}. Prior studies have examined the policies and agreements formed by user communities on decentralized social media platforms like \texttt{mastodon.social} and Bluesky \citep{gehl_digitalcovenant_2023, tosch_privacypolicies_2024,nicholson_mastodonrules_2023}, as well as documented governance challenges faced by users in administrative roles \citep{anaobi_willadmins_2023,zhang_troubleparadise_2024,hassan_exploringcontent_2021}, with recurring concerns around re-concentration of power \citep{huang_decentralizedsocial_2024,raman_challengesdecentralised_2019} at key points of decision-making \citep{jhaver_decentralizingplatform_2023}.
Crucially, the \textit{protocols}---which define rules governing the exchange of data across communication channels---undergirding decentralized social media serve as the foundation for practices in these systems. They can vary in ways that reflect protocol developers' goals and values, operationalizing and envisioning ``decentralization'' in distinct ways. As such, concerns about power in decentralized social media must also be examined in relation to the politics embedded in each protocol. In the absence of such context, researchers risk overgeneralizing user behavior and misattributing the sources and significance of key challenges in decentralized social media. Despite this, there has been a relative lack of scholarly attention to protocols themselves. 

We aim to offer a lens through which social computing scholars and developers can see the politics of decentralized social media protocols. Assessing protocols can be challenging because they are abstracted sets of rules that most users do not interact with directly. This can make it difficult to understand how values underpinning protocols translate into actual systems. Additionally, each protocol may use different terms for the same concept, complicating efforts to compare and evaluate protocols toward identifying improvements for protocol design at large. Recent empirical work on decentralized social media has predominantly focused on specific apps that only reflect a portion of their respective decentralized social media protocol, with most research looking at the case of \texttt{mastodon.social} (one of the many platforms built using the ActivityPub protocol, and described as an alternative to $\mathbb{X}$) \citep[e.g.,][]{bono_explorationdecentralized_2024,lee_usesgratifications_2023,cava_driverssocial_2023}. 
We present a novel conceptual framework that offers a consistent lexicon of the core socio-technical components that any feed-based decentralized social media protocol must define, enabling description and comparison of protocols as they are operationalized. 
Our framework components are categorized into two levels: architectural components, which represent a set of crucial technical infrastructure; and interaction components, which represent a set of social and technical functionalities that shape user experience. 

After providing an overview of the framework, we apply it to describe four major decentralized social media protocols, showing how each protocol takes a unique approach to decentralization. Drawing on protocol documentation, media coverage, and first-hand interviews with protocol developers and experts, we examine how the unique arrangement of socio-technical components outlined by each protocol reflects the particular goals and priorities of the developers, who appear to make different trade-offs accordingly. In particular, we discuss how control over key components becomes distributed differently across the protocols, allocating decision-making power over questions of identity, curation, and infrastructure. We argue that examining protocols as socio-technical artifacts can help identify opportunities to not only revise protocols themselves but develop tools that help address concerns, i.e., those related to power.

The contribution of this work is three-fold. 
First, we present the aforementioned framework, which can be used to understand the socio-technical artifacts produced by decentralized social media protocols beyond a specific extant case. 
Second, we provide a grounded description of four major decentralized protocols that would be of interest to scholars aiming to familiarize themselves with this emergent area, with a focus on making their distinct politics salient. 
Third, we take advantage of the consistency of language to identify three key issues apparent \textit{across} the four protocols that would be of interest for future work: power over identity, power over curation, and power over infrastructure. We call for a greater focus on protocol design in CSCW and social computing research, ending with a discussion of how the framework might catalyze future work given the challenges to decentralization highlighted by these issues.

\section{Background}
\subsection{Shaping values in the design of social media} \label{bg_values}
Scholars have long recognized that the design choices underpinning social computing technologies can embody social, political, and cultural values in consequential ways. As Langdon Winner \citep{winner_artifactshave_1980} observed, artifacts have politics: they reflect and reinforce particular power dynamics, social norms, and institutional priorities. Prior work examining values in the design of social media has paid special attention to the politics of \textit{platforms} \citep[e.g.,][]{gillespie_politicsplatforms_2010,christin_internalfractures_2024,duffy_platformgovernance_2023}. Recognizing that the term ``platform'' has particular discursive power \citep{gillespie_politicsplatforms_2010}, we use the term to refer to a digital system that mediates interaction and content exchange between users \citep{gillespie_regulationplatforms_2018,gorwa_whatplatform_2019}. We focus on \textit{feed-based} social media platforms, i.e., those that center feeds as a central interaction mechanism, because of the dominance and widespread use of such platforms in the past 15 years. Research examining values in feed-based social media typically considers specific features and design choices of centralized, corporate-owned social media platforms \citep[e.g.,][]{caplan_tieredgovernance_2020,duffy_platformgovernance_2023,crawford_whatflag_2016}. For example, \citet{crawford_whatflag_2016} argues that user ``flagging'' tools give platforms rhetorical cover for moderation decisions, while concealing the underlying negotiations around content removal. Studies show how algorithmic feeds and moderation filters not only play a substantial role in user interaction and behavior \citep{lustig_algorithmicauthority_2016,caplan_isomorphismalgorithms_2018,duffy_platformgovernance_2023,massanari_gamergatefappening_2017}, but can also perpetuate societal biases (e.g., concerning social identity groups) \citep{diasoliva_fightinghate_2021}. 

These bodies of work raise the question of \textit{who} can set the values that guide platform design and \textit{which} values are thus reflected in social media, often emphasizing subsequent harms and consequences (e.g., false positives in moderation). In an attempt to shift the values shaping our social media experiences, researchers and designers have worked to develop and test novel end-user tools that allow people to gain more control over their experiences~\citep{jhaver_personalizingcontent_2023}, as well as in support of their peers~\citep{mahar_squadboxtool_2018,jhaver_onlineharassment_2018} and their community \citep{jhaver_designingword_2022,jhaver_humanmachinecollaboration_2019,zhang_policykitbuilding_2020}. This work has been inspired by value-sensitive design \citep{friedman_valuesensitivedesign_1996} approaches, as well as guided by greater awareness of the experiences of vulnerable and marginalized populations (who face greater risks of online harm \citep[see][]{vitak_identifyingwomens_2017,schoenebeck_onlineharassment_2023}). However, most contemporary feed-based social media platforms---like $\mathbb{X}$ (formerly Twitter), Facebook, or TikTok---are proprietary systems that operate in a centralized manner. They are often closed and opaque, meaning end-user tools offer only limited recourse or can be easily undermined by platform-level decisions (e.g., limiting an API) \citep{peters_howreddit_2023}.

Crucially, social media platforms need not be fundamentally closed or centralized~\citep{jhaver_decentralizingplatform_2023}. They can be open-source, meaning that anyone can in theory use, change, and share them \citep{demonnink_networkplatform_2024}. They can also be decentralized, technically distributing decision-making around key functionalities---such as data storage, content moderation, and identity management---across a network of independent servers managed by different kinds of actors, such as everyday users \citep{tosch_privacypolicies_2024,hwang_trustfriction_2025}. \citet{masnick_protocolsnot_2019} lays out this alternative vision for social media, calling for a focus on developing social media \textit{protocols}, not platforms. A protocol is a set of rules governing the exchange of data between defined communications channels. By creating a shared, open standard, protocols can be hugely impactful and have played a major role in modern digital technologies: the early Internet was centered around protocols that still form the backbone of many digital applications and services today, such as SMTP (Simple Mail Transfer Protocol, which enables email) and HTTP (Hyper Text Transfer Protocol, which undergirds the World Wide Web) \citep{masnick_protocolsnot_2019}.

In the context of social media specifically, protocols lead to platforms but not all platforms have protocols. Developers and engineers build a system---which, if protocol-based, implements a protocol into a technical artifact---and the deployment of the system results in what we recognize as social media platforms. 
By standardizing and publishing rules about data exchange, protocols enable open, decentralized, and interoperable systems of digital tools and services (described in greater detail in \S \ref{bg_decentralization})---these distinguish social media platforms built on protocols from centralized, corporate-operated ones that have been the focus of much prior research. 

Protocols are inherently normative, and they are thus consequential because they can have normative consequences for the social and technical systems built on top of them. A social media protocol may establish rules about how user information is accessed and by whom, which actors can determine content moderation approaches (e.g., algorithmic filtering, whether there is room for dispute, and so on), or whether third parties may develop and connect apps through an API. Seen otherwise, protocols allocate power over decision-making in how social media platforms mediate interaction. Thus, focusing on the design and implementation of decentralized social media protocols can help surface and create opportunities to determine the values that undergird social media platforms. Further, when decentralized social media protocols are instantiated as a complex assemblage of software, they become distinct artifacts whose politics can be analyzed.

\subsection{Re-imagining social media with decentralized protocols} \label{bg_decentralization}
Recent years have seen significant growth in decentralized social media protocols as viable alternatives to mainstream centralized platforms. The most prominent example of this is the surge of users on Bluesky,\footnote{\texttt{\href{https://bsky.app/}{https://bsky.app/}}} a decentralized social media platform using the AT Protocol, following the election of Donald Trump in November of 2024, credited in large part to the dissemination of right-wing rhetoric on platforms like $\mathbb{X}$ leading up to the election \citep{russell_competitorbluesky_2025, boran_elonmusks_2024}. People especially discontent with Elon Musk's ownership of $\mathbb{X}$ sought out decentralized, non-corporate alternatives for microblogging that were perceived as safer and more trustworthy \citep{boran_elonmusks_2024, jeong_exploringplatform_2024, brice_howstop_2024}.

Because many recent decentralized social media protocols have emerged in response to persistent concerns around trust, safety, and power imbalances on centralized platforms \citep[e.g.,][]{schoenebeck_onlineharassment_2023,chancellor_thyghgappinstagram_2016,haimson_transtime_2020}, we examine protocols that explicitly frame their technical \textit{decentralization} as a solution to these issues.\footnote{The exact protocols examined are discussed in \S \ref{s_scope}} Scholars, activists, and developers have increasingly advocated for reimagining social media systems with decentralization at their core, emphasizing the potential to foster ecosystems that better serve the public interest—particularly in terms of safety, equity, and inclusivity \citep[see][]{cohn_fediversecould_2022,karppi_ifnot_2023,masnick_protocolsnot_2019}. These arguments tend to follow two main lines of reasoning.

First, decentralization disperses power away from a single actor. Most popular social media systems today are centralized in design, concentrating control over data, algorithms, and interactions to a single actor---historically, a company, which is more likely to prioritize corporate interests over the well-being of its user base. For example, popular platforms like Instagram and LinkedIn train their internal large language models with user data for new generative AI features that attract more users but raise serious privacy concerns, such as the leaking of user information \citep{brice_howstop_2024, mauran_metausing_2024}. In contrast, decentralized social media protocols like ActivityPub operationalize systems designed to distribute decision-making---such as how user data is accessed and used---across a broader group of people. For example, the developers behind Pixelfed---a social photo-sharing software implementation of the ActivityPub protocol---maintain their own platform called \texttt{pixelfed.social},\footnote{\texttt{\href{https://pixelfed.social}{https://pixelfed.social}}} but also allow individuals to create their own independent instances of the platform. These instances operate autonomously but can still interconnect with others. Under this model, governance decisions are typically made by instance administrators, who are usually more accessible and approachable for users to communicate with directly (i.e., relative to the CEOs of large companies operating social media platforms) \cite{anaobi_improvingcontent_2024, anaobi_willadmins_2023, hwang_trustfriction_2025, zhang_troubleparadise_2024}. 

Second, decentralized social media protocols typically encourage actors at different points of decision-making to build or adapt interoperable technologies. This allows for the development of technologies that are well-aligned to unique needs, unlike the one-size-fits-all model imposed by centralized social media. For example, the Farcaster protocol allows a wide range of specialized apps to operate independently while remaining interconnected within the same ecosystem. Platforms built using Farcaster like Eventcaster\footnote{\texttt{\href{https://events.xyz/}{https://events.xyz/}}} (comparable to event ticketing platform Eventbrite) or Launchcaster\footnote{\texttt{\href{https://launchcaster.xyz/}{https://launchcaster.xyz/}}} (comparable to the tech-discovery platform Product Hunt) prioritize custom features that drive community-building by tending to the needs of their users. In this sense, decentralization encourages innovation in the tools and technologies built for social media \citep{keller_futureplatform_2021,masnick_protocolsnot_2019}. Masnick \citep{masnick_protocolsnot_2019} argues that whereas inter-platform competition (i.e., between platforms like Facebook or $\mathbb{X}$) is limited, decentralized social media protocols could encourage a new level of competition within one social media ecosystem as ``anyone could present a new interface, or new features'' as well as ``create entirely new areas for innovation, including in ancillary services, such as parties that focus on providing better content moderation tools or the competing databanks.'' Theoretically, this would make social media ecosystems more responsive and adaptable to emerging issues and harms. 

\subsection{Assessing the promises and challenges of decentralized social media}  \label{bg_challenges}
Although presented as a solution to the problems we observe in social media, decentralized social media faces challenges too. Many of these echo persistent problems around online behavior such as dealing with spam and trolling. Because decentralization often shifts governance and moderation responsibilities to individual nodes or communities within a broader network \citep{gehl_digitalcovenant_2023}, prior work has explored how decentralized social media might address issues like rule-breaking through locally defined mechanisms—such as community-specific guidelines \citep{nicholson_mastodonrules_2023} or the ability for different nodes to sever ties with one another across shared protocols \citep{colglazier_effectsgroup_2024}. Framing decentralized social media as an emergent empirical site, these studies examine user behavior in light of the fact that more individuals now have greater agency as admins of their own platforms. Findings echo some of the dynamics seen in other online community contexts such as Reddit or Wikia. For example, prior research examines how admins manage the privacy policies of their servers, finding that admins very infrequently tailor privacy policies---a pattern of isomorphism seen in other types of online communities \citep{kiene_identitylegitimacy_2024}. Related studies focus on identifying challenges faced by users acting as administrators \citep{zhang_troubleparadise_2024}, as well as developing tooling to address them \citep{anaobi_improvingcontent_2024,agarwal_decentralisedmoderation_2024a}.

Another set of work focuses on the potential for re-centralization in decentralized social media. This includes scholarship analyzing the ethos of decentralized social media \citep{demonnink_networkplatform_2024,gehl_digitalcovenant_2023}. Research notes that decentralized social media sees user-driven, infrastructure-driven, and content-driven pressures toward centralization due to issues related to costs and user experience \citep{raman_challengesdecentralised_2019}. \citet{tosch_privacypolicies_2024} suggest that prior norms about consent, such as agreeing to privacy policies or terms of service, are holdovers from centralized platforms but poor fits for decentralized models. \citet{hwang_trustfriction_2025} further outlines how decentralization leads to unique issues of coordination that must be addressed in future design work in decentralized social media.

The persistent and unique challenges documented in these studies highlight how designing a decentralized social media protocol is neither straightforward nor obvious. Just as social media can embody centralized or decentralized principles, decentralized social media protocols may envision and operationalize ``decentralization'' differently. Notably, much of the prior empirical work mentioned above focuses on Mastodon platforms that use the ActivityPub protocol. However, many other protocols exist, made by developers who make distinct trade-offs in design choices that align with their specific values, priorities, and goals. Understanding the different instantiations of ``decentralization'' is crucial for identifying the pros and cons of each approach, which can, in turn, inform the future development of decentralized social media more broadly. A key challenge in evaluating social media protocols to this end is the lack of a consistent way to understand how the values motivating protocol design translate into concrete socio-technical systems. In this work, we present a framework to directly address this gap, offering a lexicon of socio-technical components that allows us to describe and compare decentralized social media systems, toward evaluating the values they encode. As decentralized social media moves from the margins to the mainstream, this kind of analysis becomes urgent: today's choices in protocol design could determine who holds power, who gets heard, and what futures become possible online.

\section{Method}
\subsection{Real-world Protocol Overview} \label{s_scope}
Although a myriad of decentralized social media protocols exist, we limit our analysis to four that have gained significant mainstream traction. These protocols represent a substantial part of the ecosystem in terms of adoption and developer activity, making them both theoretically relevant and practically impactful.

\subsubsection{ActivityPub}
ActivityPub describes itself as a ``decentralized social networking protocol [that] provides a client-to-server API for creating, updating and deleting content, as well as a federated server-to-server API for delivering notifications and content'' \cite{lemmer-webber_activitypubw3c_2018}. By enabling a network of servers individually operated by users, ActivityPub aims to give users more control over their data and interactions through decentralization. The protocol first emerged in 2018 under the auspices of the World Wide Web Consortium (W3C) as a means to return the web to a time ``before everything got locked down into a handful of walled gardens'' \cite{lemmer-webber_activitypubrocks_2021}. \citet{gehl_activitypubnonstandard_2023} notes the lack of corporate social media involvement in the development of ActivityPub; its initial developers---self-identified members of the LGBTQ+ community---sought to build spaces free from the harassment they experienced on mainstream social media such as Twitter \citep{karppi_ifnot_2023}, and in doing so also wanted the protocol to also be largely free and open source. One of the best-known implementations of ActivityPub is Mastodon, along with its flagship platform \texttt{mastodon.social}, which saw a surge of interest in the wake of Twitter's acquisition in 2022 \citep{boran_elonmusks_2024}. Many other social media developers have also been drawn to ActivityPub because of its technical intuitiveness and the interoperability of its platforms, including those behind Flipboard, Threads (by Meta), and Truth Social \cite{sherr_twitterwho_2023, backlinko_numberthreads_2024, cohen_whattruth_2022}.

\subsubsection{AT Protocol}
The Authenticated Transfer (AT) Protocol describes itself as innovating upon ``existing data models from the (decentralized social) protocol family'' \citep{atprotocol_protocolspecs_2024}. A defining characteristic of AT Protocol is the goal to replicate traditional social media experiences so that new users can more easily transition to decentralized social media. AT Protocol was initially developed by a research team at $\mathbb{X}$ (then, still Twitter) investigating how the platform could be better decentralized, before separating from the centralized platform during $\mathbb{X}$'s acquisition \citep{russell_competitorbluesky_2025}. The team then developed into its own corporate entity (Bluesky Social PBC) and sought funding from venture capital. Formally unveiled in 2022 with the launch of Bluesky \citep{robertson_twittersopensource_2022}, the flagship social media platform on the protocol, AT Protocol seeks to be a ``networking technology created to power the next generation of social applications'' \citep{atprotocol_protocolhomepage_2024}. It aims to facilitate the creation of social media systems that are ``usable, scalable, portable, and accountable to (their) users'' \citep{ipfs_blueskyipld_2022}, including connections between third-party clients (e.g., Graysky\footnote{\texttt{\href{https://graysky.app/}{https://graysky.app/}}}, Klearsky\footnote{\texttt{\href{https://klearsky.pages.dev/}{https://klearsky.pages.dev/}}}) using the protocol but providing different interfaces and features. AT Protocol is thus meant to enable customizable user experiences without sacrificing connectivity. In late 2024, membership on the Bluesky platform grew rapidly, reaching over 25 million users \citep{schwarz_websitetracks_2024}. 

\subsubsection{Nostr}
Nostr (Notes and Other Stuff Transmitted by Relays) describes itself as a protocol ``designed for simplicity, that aims to create a censorship-resistant global social network'' \cite{nostr_nostrsimple_2024}. The development of Nostr was led by an open-source developer who goes by the moniker @fiatjaf, with initial financial support from Jack Dorsey\footnote{Former CEO of Twitter, now $\mathbb{X}$.} \citep{delcastillo_meetfiatjaf_2023}. The protocol---which relies largely on open-source contribution for its upkeep---launched in 2020 to offer users a platform where they can express themselves without fear of deplatforming or algorithmic manipulation associated with traditional social media \cite{bitcoinmagazine_nostrprotocol_2023}. For example, Damus,\footnote{\texttt{\href{https://damus.io/}{https://damus.io/}}} one of the largest Nostr platforms \citep{torpey_hereswhy_2023}, markets itself as a bastion of free speech; it takes pride in moments when governments ban its use, as an indication of its value and impact \citep{torpey_hereswhy_2023,liao_damuspulled_2023}. Nostr's ties to cryptocurrencies and blockchain has made it popular among Bitcoin enthusiasts, particularly those valuing privacy and security, user autonomy, and decentralized control. Nostr utilizes cryptographic keys for user authentication and identification to enable secure content-sharing (akin to how Bitcoin wallets work) \cite{nostr_nostrsimple_2024}. Another popular platform using Nostr, Primal\footnote{\texttt{\href{https://primal.net/}{https://primal.net/}}} initially started as a Bitcoin wallet before evolving into a social media platform (to try to improve user retention) \cite{destries_primalsecures_2023}. 

\subsubsection{Farcaster}
Farcaster describes itself as a ``sufficiently decentralized social network built on Ethereum'' where users ``own their accounts and relationships with other users and are free to move between different apps'' \citep{farcaster_gettingstarted_2024}. The protocol was launched in 2020 by a group of prominent blockchain engineers and a large backing from venture capital firms like a16z  \citep{silberling_farcastercryptobased_2024}. The creators were motivated by the belief that $\mathbb{X}$ had become a ``slow-moving company without much innovation,'' creating a market opportunity for a ``fundamentally new, exciting thing that is a protocol and decentralized'' \cite{gabriele_futurefarcaster_2024}. Farcaster also has strong ties to the cryptocurrency community, specifically enthusiasts of Ethereum, as it integrates a scaling solution on top of the Ethereum blockchain\footnote{\texttt{\href{https://etherscan.io/}{https://etherscan.io/}}} to store certain user data. This allows users to create accounts on Farcaster platforms by connecting their crypto wallets, making their accounts independent of any social media service. Social media platforms built using Farcaster are not entirely blockchain-based, however. Farcaster seeks to leverage both the verifiability and immutability of blockchain technologies ``when security and consistency are critical'' and the utility of traditional servers ``when performance and cost are critical'' \cite{farcaster_architecturefarcaster_2024}. Warpcast,\footnote{\texttt{\href{https://warpcast.com/}{https://warpcast.com/}}} the flagship Farcaster platform, exemplifies blockchain integration by requiring users to join via a wallet-based buy-in fee.

\subsection{Inductive analysis}
We took an iterative, inductive approach to devise a framework (\S \ref{s_framework}) for ``seeing'' the politics of social media protocols. We began by gathering publicly available resources about each protocol, including documentation, white papers, and interviews from media outlets like The Verge, WIRED, Forbes, and TechCrunch. We also used various social media platforms built using the four protocols of interest first-hand, gaining ethnographic-style insights into the nuanced ways that protocol-level design decisions shaped what end users could do, see, and govern. For example, some platforms built with specific protocols provided explicit mechanisms for moderation, while others leaned heavily on user-driven reporting and filtering tools. Our goal was to define a consistent set of components to support comparative analysis of decentralized social media protocols.

Drawing from the aforementioned sources, the first author began by generating and refining an initial list of components intuitively deemed essential for all feed-based social media platforms, including (but not limited to): posts, likes, replies, profiles, moderation, feeds, and other core functionalities. The research team continuously met weekly over Zoom and communicated asynchronously over Slack to discuss and iterate on this list, identifying components that were missing (e.g., protocol-defined affordances for identity) and clustering granular components into broader categories (e.g., ``posts'', ``quotes'', and ``reactions'', collectively becoming ``\Act s'') through discussion, re-review of protocol documentation, and stress testing the viability of the framework to describe existing protocols. As we did so, we contextualized our terminology with sources that signaled the political, philosophical, and social values motivating their development.

\begin{table}[t]
    \centering
    \begin{tabular}{lll}
        \toprule
        Alias & Archetype & Affiliation \\
        \midrule
        P1 & Builder & ActivityPub, AT Protocol \\ \hline
        P2 & Builder & AT Protocol, ActivityPub \\ \hline
        P3 & Builder & Nostr, ActivityPub \\ \hline
        P4 & Generalist & ActivityPub, AT Protocol, Nostr, Farcaster \\ \hline
        P5 & Generalist & ActivityPub, AT Protocol, Farcaster \\ \hline 
        P6 & Creator & ActivityPub \\ \hline
        P7 & Builder & Farcaster, Nostr \\ \hline
        P8 & Generalist & ActivityPub, AT Protocol, Nostr, Farcaster \\ \hline
        P9 & Creator & ActivityPub, AT Protocol  \\ \hline
        P10 & Generalist & ActivityPub, AT Protocol, Nostr, Farcaster  \\ \hline
        \bottomrule
    \end{tabular}
    \caption{List of interviewees and their decentralized protocol experience.}
    \label{tab:sample}
    \vspace{-5mm}
\end{table}

\subsection{Triangulation}
\noindent To evaluate how well our framework captured the socio-technical nuances of decentralized social media protocols, we interviewed ten individuals deeply embedded in these ecosystems from September 2024 to April 2025. Interviews, conducted via Zoom and averaging 47 minutes, were structured around three distinct archetypes to capture diverse perspectives:
\begin{itemize}
    \item \textbf{Creators:} People involved with conceiving a protocol, contributing to the initial codebase, and defining foundational architecture (e.g., ActivityPub's founding developers).
    \item \textbf{Builders:} Individuals implementing platforms built using a protocol, such as Graysky for AT Protocol or Primal for Nostr. 
    They provided valuable insights into how protocol conventions influence user experience and platform functionality.
    \item \textbf{Generalists:} Experts with knowledge of multiple protocols. They offered breadth over depth, providing comparative insights to critique and verify our categorization analyses.
\end{itemize}

Interviews allowed us to iteratively refine the framework by triangulating our user experiences, public documentation, and developer insights. Participants received a draft of the framework in advance and read it aloud during sessions, pausing to offer feedback, corrections, and clarifications. This process surfaced misconceptions and nuanced details about protocol design. Discussions with creators and builders focused on their areas of expertise, with creators sharing ideological motivations and design trade-offs and builders highlighting protocol features and implementation challenges. Generalists helped us better compare protocols, given their cross-protocol experience.

\newpage 

\section{A Framework for Seeing Feed-Based Social Media and their Protocols} \label{s_framework}
We introduce a framework to provide a consistent and accessible lexicon through which we can describe feed-based decentralized social media protocols. The framework's four \textit{architectural components} represent the technical infrastructure necessary for social media platforms to function; its six \textit{interaction components} are the socio-technical functionalities end users interface with.

\subsection{Architectural Components}
\begin{figure}[h!]
    \centering
    \hspace{-20mm}\subfloat[
        \parbox{\textwidth}{A centralized social media system using the framework's lexicon.}]{
        \hspace{20mm}\includegraphics[width=0.30\textwidth]{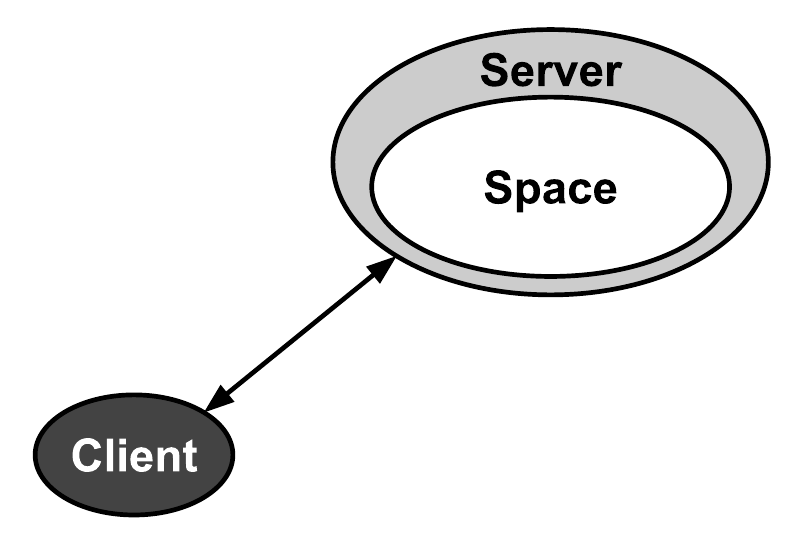} 
        \label{fig:architectural_components_centralized}

    } \\[1em]
    
    \subfloat[ActivityPub]{
        \includegraphics[width=0.475\textwidth]{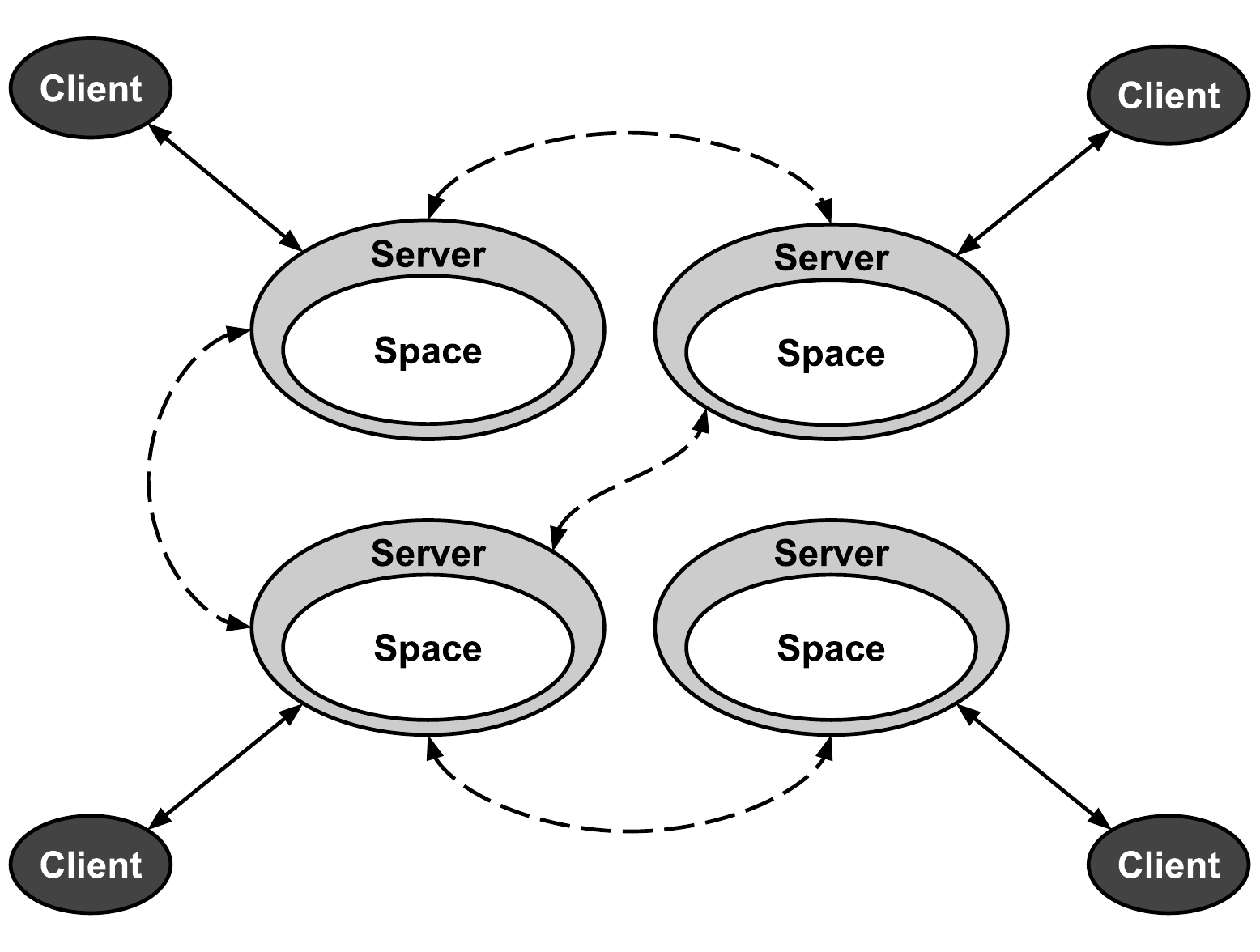}
        \label{fig:architectural_components_activitypub}
    }
    \hfill
    \subfloat[AT Protocol]{
        \includegraphics[width=0.475\textwidth]{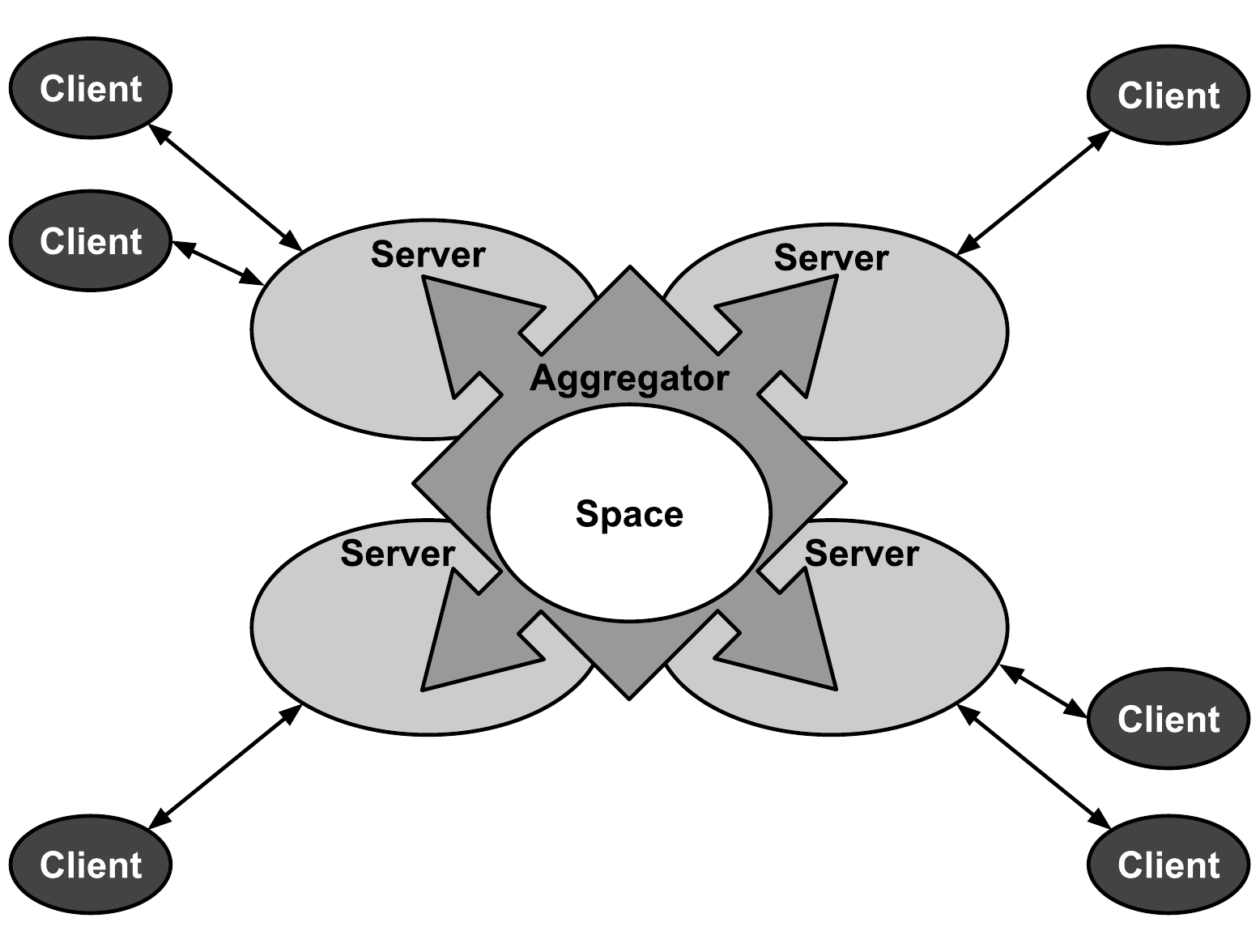}         \label{fig:architectural_components_atproto}
    } \\[1em]
    
    \subfloat[Nostr]{
        \includegraphics[width=0.475\textwidth]{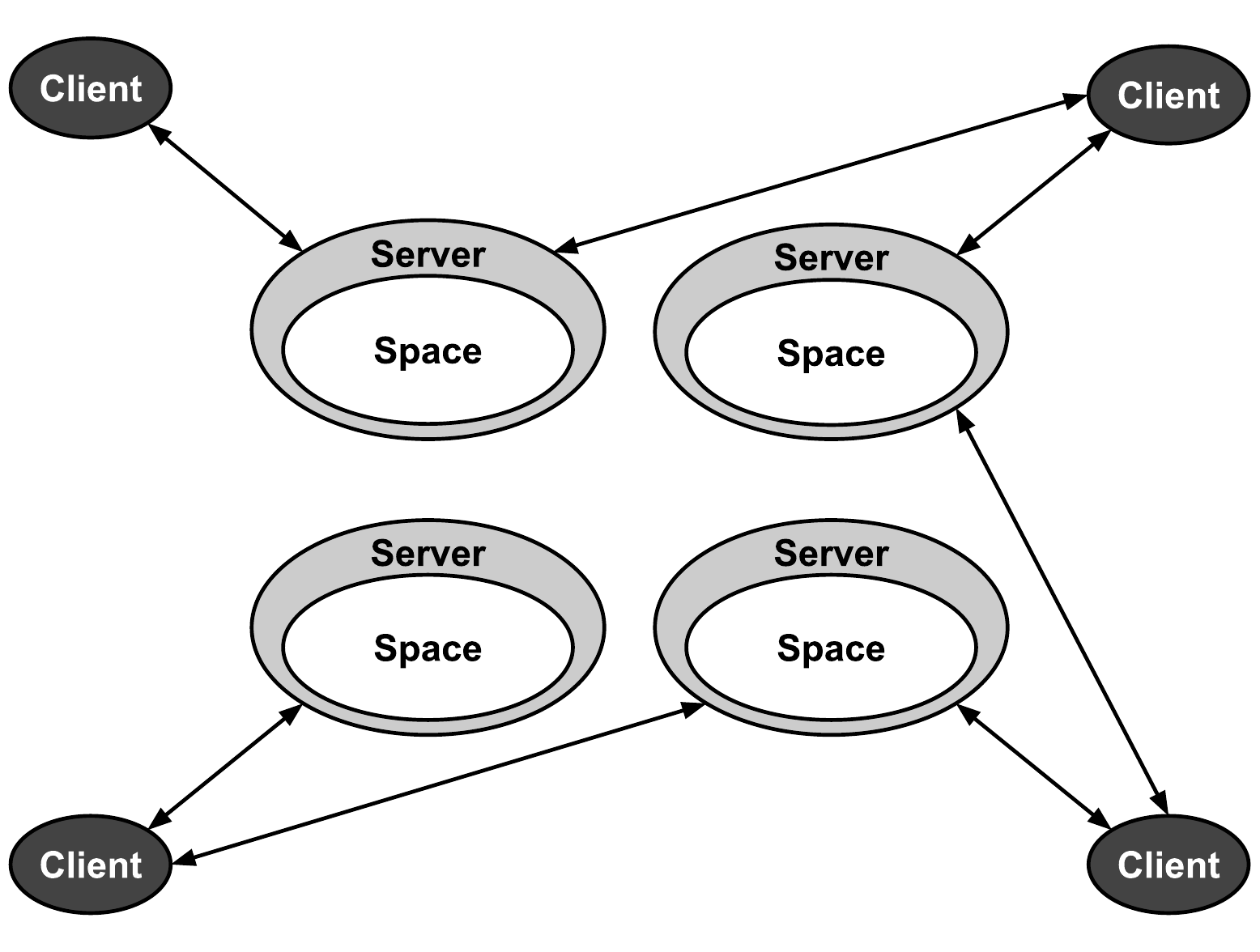} 
        \label{fig:architectural_components_nostr}
    }
    \hfill
    \subfloat[Farcaster]{
        \includegraphics[width=0.475\textwidth]{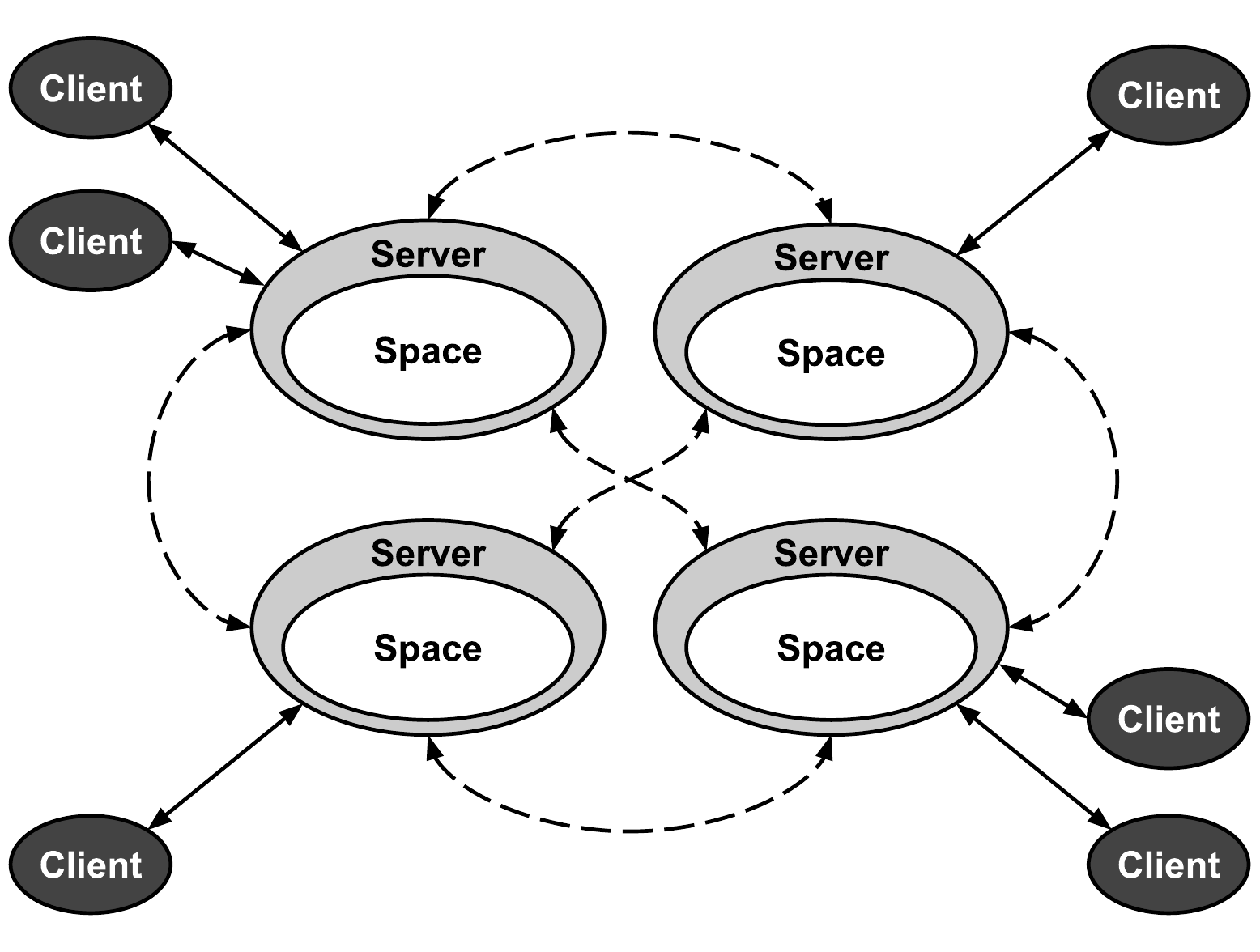} 
        \label{fig:architectural_components_farcaster}
    }
    \caption{Architectural component layouts for our focus protocols, and centralized social media for comparison.}
    \label{fig:architectural_components}

\end{figure}

\newpage

\noindent Architectural components are the technical building blocks that social media protocols define rules for when creating social platforms. We define them as follows:

\begin{itemize}
    \item \textbf{\Cli s} are the front-end software used to interact with others via the conventions defined by a protocol. They serve as the gateway (and interface) through which users interact with others, i.e., send and receive content. A \Cli\ is owned by whoever creates or maintains it.
    \item \textbf{\Spa s} are the virtual environments where users engage with one another. When users sign up to join a social media platform (usually via a \Cli), they become a part of a \Spa. A \Spa\ is owned by whoever creates or maintains it by having full access to its Servers.
    \item \textbf{\Ser s} are where back-end functionality occurs and where data for Spaces are stored. We define a `\Ser' as a single computer or a group of computers working in tandem owned by the same entity or organization. Decentralized systems typically have many \Ser, maintained by different individuals and distributed across several locations. A \Ser\ is owned by whoever creates or maintains it.
    \item \textbf{\Agg s} are computing units that collect content from multiple \Ser, compile it into a single ledger, and distribute the consolidated version back to all \Ser\ in their network.
\end{itemize}

\noindent Together, \Ser s, \Cli s, and \Spa s comprise the basic technical infrastructure that functions as social media platforms, while \Agg s are occasionally present to optimize connectivity between Servers. In a typical centralized platform (see Figure \ref{fig:architectural_components} (a)), only three of the four architectural components are relevant: one \Cli\ gateway to one \Spa, which is kept within one \Ser. Because there is only one \Ser\ (owned by the centralized entity), \Agg s are unnecessary.

\begin{table}[t]
    \centering
    \begin{tabular}{|p{2cm}|p{2.5cm}|p{2.5cm}|p{2.5cm}|p{2.5cm}|}
    \hline
    \textbf{} & \textbf{ActivityPub}& \textbf{AT Protocol}& \textbf{Nostr} & \textbf{Farcaster}\\ 
    \hline
    \textbf{\textsc{{Client}}} & Client & App & Client & App \\ 
    \hline
    \textbf{\Spa} & Instance & Atmosphere & \textit{No Specific Name} &  \textit{No Specific Name} \\ 
    \hline
    \textbf{\Ser} & Instance & \raggedright Personal Data Server (PDS) & Nostr Relay & Hub \\ 
    \hline
    \textbf{\Agg} & N/A & \raggedright AT Protocol Relay & N/A & N/A \\ 
    \hline
    \end{tabular}
    \caption{Semantics of architectural components across our four decentralized protocols. Note that Nostr and Farcaster have \Spa-like concepts without specific names. ActivityPub, Nostr, and Farcaster do not include Aggregators.}
    \label{tab:architectural_components_comparison}
\end{table}

\newpage

\subsection{Interaction Components}

\begin{figure}[t]
    \centering
    \includegraphics[height=150pt]{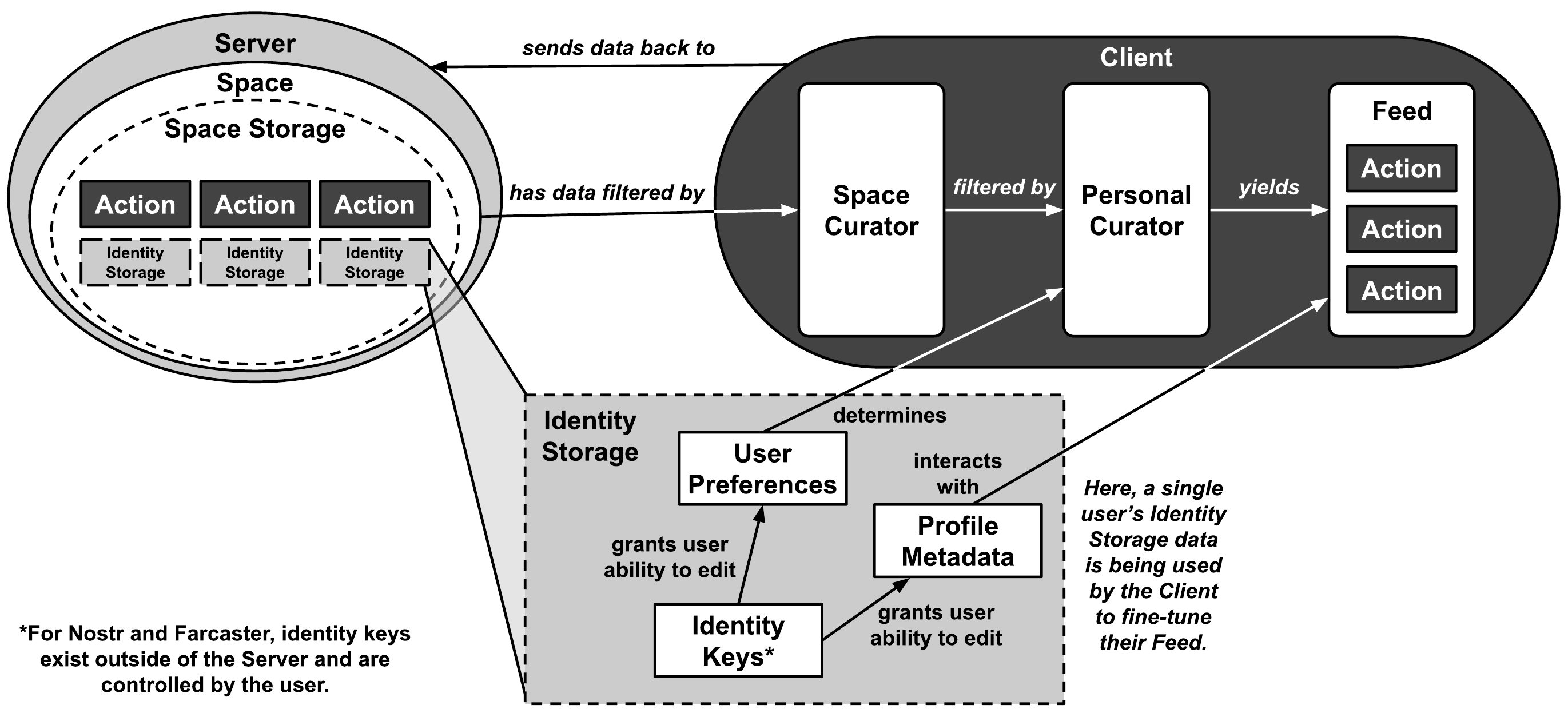}
    \caption{Interaction components for a generic centralized feed-based social media platform.}
    \label{fig:interaction_components_centralized}
\end{figure}

\noindent The \textit{interaction components} are the social and technical functionalities that guide how users communicate, share content, and engage with one another. Architectural components often influence the design of interaction components.

\begin{itemize}
    \item \textbf{\Act s} encompass every type of interaction that users can invoke on a social media platform, such as posting content, following others, or reacting to others' content. \textsc{Actions} are stored as data in a \Ser, and a \Cli's design usually dictates how \textsc{Actions} work.
    \item \textbf{\Fee s} are the interfaces within \Cli s where users engage with an ordered set of \Act.
    \item \textbf{\SSt} refers to the database(s) where \Act s are stored. It constitutes the content of a \Spa, and is usually kept in the \Ser s that underpin a given \Spa.
    \item \textbf{\ISt} refers to the data associated with a user's identity, including usernames, avatars, and other profile metadata, and their preferences, such as following or block list data. This type of data is always a subset of \SSt, but we highlight \ISt\ separately because for some protocols, identity keys---which are essential for enabling users to edit their preferences and profile metadata---are managed by the \Ser\ owner as part of \ISt. In other cases, identity keys are held directly by the user, shifting responsibility from the \Ser\ to the individual. Regardless of location, these keys are essential for user control over identity and preferences.
    \item \textbf{\SCu s} are the tools, systems, and individuals responsible for maintaining the quality and appropriateness of content within a \Spa. These can work at both the \Cli\ and \Ser\ levels, e.g., moderation teams and content filtering algorithms. This can also include composable moderation, where individual \Cli-level preset filters---maintained by various actors---can be combined to create a flexible, customizable moderation strategy.
    \item \textbf{\PCu s} are features and mechanisms within the \Cli\ that allow individual users to manage their own experience---such as choosing whom to follow, block, or mute. While conceptually distinct from \SCu s, their behavior is closely guided by the preferences recorded in \ISt. This creates a clear division of roles: \ISt\ serves as the persistent record of a user's personal preferences and identity-related settings, while \PCu s are the functional components that interpret and act on those preferences within the user interface. Together, they enable users to shape a personalized and consistent experience.
\end{itemize}

\noindent In a typical centralized social media platform (see Figure \ref{fig:interaction_components_centralized}), when a user joins a \Spa\ via a \Cli, they gain access to their \ISt. This is a subset of the broader \SSt\ where a user's preferences and profile metadata are stored. While data flows from the singular \Ser\ to the \Cli, the \Cli's algorithmic \Fee\ largely determines the user's experience within the \Spa. Content shown in the \Fee\ is typically filtered first by a centralized \SCu\ that enforces platform-wide rules, then by \PCu s, such as user-defined blocks or mutes. 

\subsection{Describing Decentralized Social Media Protocols}

\begin{table}[t]
    \centering
    \begin{tabular}{|P{2cm}|P{2.5cm}|P{2.5cm}|P{2.5cm}|P{2.5cm}|}
    \hline
    \textbf{} & \textbf{ActivityPub} & \textbf{AT Protocol} & \textbf{Nostr} & \textbf{Farcaster}\\
    \hline
    \textbf{\textsc{Actions}} & Activity & Record & Event & Cast \\ 
    \hline
    \textbf{\Fee} & Inbox & Feed & \textit{No Specific Name} & Feed \\ 
    \hline
    \textbf{\SSt} & Instance Database & PDS Database & Nostr Relay Database & Farcaster Hub Database + Blockchain \\ 
    \hline
    \textbf{\ISt} & Table Entry Within Instance Database & Document Within PDS Database & Document Within Nostr Relay Database & Document Within Hub Database + Crypto Wallet \\ 
    \hline
    \textbf{\SCu s} & Instance Admin Moderation + User Selected Inboxes & Composable Content Moderation  + \Cli\ Specified & \textit{\Cli\ Specified} & \textit{\Cli\ Specified} \\ 
    \hline
    \textbf{\PCu s} & \multicolumn{4}{P{10.5cm}|}{Account Follows, Account Blocks, Account Mutes, \& More*} \\
    \hline
    \end{tabular}
    \caption{Semantics of interaction components across our four decentralized protocols. Note \ISt\ is often tied to the \SSt\ infrastructure. *Additionally, across the four protocols, \PCu s are typically defined at the \Cli-level.}
    \label{tab:interaction_components_comparison}
\end{table}

Drawing from documentation, blog posts, and public media articles---and validated in our interviews with protocol developers---we now apply the framework to describe four real-world decentralized social media protocols and show how it can articulate (\S 4.3.1-4.3.4) a clear socio-technical artifact---the social media system---as defined by a protocol. By leveraging a consistent lexicon, we aim to make the similarities and differences of the protocols salient (see \ref{fig:architectural_components}, as well as Tables \ref{tab:architectural_components_comparison} and \ref{tab:interaction_components_comparison}). 

\subsubsection{ActivityPub}
ActivityPub aims to foster a diverse network of smaller, independent platforms, with interoperability facilitated by W3C-backed open standards \citep{gehl_activitypubnonstandard_2023,lemmer-webber_activitypubrocks_2021,prodromou_activitypubprogramming_2024}. \Ser s---known as \textbf{``instances''}---are independently operated and can selectively communicate with other \Ser s, directly sharing and receiving \Act s from and to their respective \SSt\ as shown in Figure \ref{fig:architectural_components_activitypub} \cite{lemmer-webber_activitypubw3c_2018}. Each \Ser\ can define its own \SCu s, such as moderation tools and blocklists \citep[e.g.,][]{anaobi_improvingcontent_2024, zhang_troubleparadise_2024}, and users interact through a corresponding \Cli\ that surfaces \Fee s curated by both platform-wide filters and user-specific \PCu s \citep{liu_understandingdecentralized_2025,silberling_beginnersguide_2023}. \ISt\ is typically housed within the same \Ser\ as the broader \SSt, though some implementations of the protocol---like \texttt{mastodon.social} or \texttt{pixelfed.social}---support account migration across instances. Because identity key management is usually handled entirely by the \Ser, users generally only need to remember a username and password. Pre-packaged, standardized software---such as Mastodon or Pixelfed---bundle \Cli s and \Ser s, making it easy for people to launch their own \Spa s with everything working out-of-the-box \cite{hering_howran_2023, mansoux_seventheses_2020}. For example, while \texttt{mastodon.social} is the flagship Mastodon platform, anyone can spin up their own \Ser\ (with a bundled \Cli) using the same software, as indicated by the thousands of independent \Ser s doing so.\footnote{As of May 4, 2025, the official Mastodon directory indicated 8.7k instances up running with the software: \texttt{https://joinmastodon.org/server}s} The first column of Tables \ref{tab:architectural_components_comparison} and \ref{tab:interaction_components_comparison} summarize architectural and interaction components. 

\subsubsection{AT Protocol}
AT Protocol aims to ensure a consistent and intuitive user experience, enabling a seamless transition on the grounds of user experience from centralized platforms \citep{russell_competitorbluesky_2025,chayka_blueskysquest_2025}. An AT Protocol \Ser\ is specifically called a \textbf{``Personal Data Server (PDS)''}, and contributes to a global \Spa\ accessible to all AT Protocol \Cli s \cite{atproto_selfhosting_2025,sanders_yourethinking_2024}. As shown by Figure \ref{fig:architectural_components_atproto}, this shared \Spa\ is coordinated by an \Agg\ known as a {``relay''}, which aggregates data from connected \Ser s into a single ledger and broadcasts it back to them \cite{newbold_replybluesky_2024,bluesky_federationarchitecture_2025}. Data flows in a cycle: from \Cli\ to \Ser, then \Ser\ to \Agg, then \Agg\ to \Ser, and finally \Ser\ back to \Cli. \SSt\ is effectively network-wide, as any content uploaded to one \Ser\ can be redistributed via the \Agg. \Act s are surfaced to users through \Fee s, but interestingly moderation here is modular: instead of top-down enforcement by a \Ser\ or \Cli\ owner, AT Protocol supports decentralized \SCu s and user-chosen \PCu s, including customizable feed filters and third-party labeling services \cite{bluesky_whyare_2023, bluesky_blueskysmoderation_2024}. For \ISt, AT Protocol manages identity at the \Ser\ level, allowing people to authenticate by using their email and password and avoiding the need for key management on the user's part \cite{atprotocol_glossaryterms_2025}. Users can also migrate accounts---along with their otherwise \Ser-held identity keys---between \Ser s via a process called ``credible exit'' \cite{atprotocol_atprotoethos_2025}. The second column of Tables \ref{tab:architectural_components_comparison} and \ref{tab:interaction_components_comparison} summarizes the architectural and interaction components of this protocol.

\subsubsection{Nostr}
Nostr aims to minimize power consolidation through a crypto-native architecture that centers \Cli s \citep{nostr_nostrsimple_2024,delcastillo_meetfiatjaf_2023,torpey_hereswhy_2023,pluja_settingnostr_2025}. \Ser s are referred to as \textbf{``relays''} \cite{neves_howsetup_2022}.\footnote{Not to be confused with AT Protocol relays (which are Aggregators).} As shown in Figure \ref{fig:architectural_components_nostr}, each \Cli\ can interface with multiple \Ser s, pulling in content from several \Spa s to provide a diversified and rich array of content. Likewise, multiple \Cli s can connect to the same \Ser, enabling shared access to its \Spa\ \cite{pluja_nostrprotocol_2025, bitcoinmagazine_nostrprotocol_2023}. Because \Ser s do not communicate directly, Nostr does not use \Agg s; instead, inter-\Spa\ communication is handled entirely at the \Cli\ level. Content lives in the \SSt\ of each \Ser, and \Act s posted there are only visible to \Cli\ s connected to that \Ser. While \Cli s surface \textsc{Feeds} using both \SCu s and user-defined \PCu s, moderation is complicated by the fact that \Cli\ and \Ser\ owners are often independent and may have different opinions about how to moderate. Nostr's identity management tied to \ISt\ follows a crypto-native model: users generate and retain their own private keys (much like Bitcoin wallets), meaning account access is fully decentralized and non-recoverable if lost \cite{torpey_hereswhy_2023}. Because Nostr lacks popular pre-packaged, standardized software that bundles the setup of \Cli s and \Ser s together, it requires a bit more technical familiarity to launch a new platform using the protocol compared to ActivityPub \cite{neves_howsetup_2022}. The third column of Tables \ref{tab:architectural_components_comparison} and \ref{tab:interaction_components_comparison} summarize the architectural and interaction components of this protocol.

\subsubsection{Farcaster}
Farcaster aims to prioritize network resilience even at the expense of accessibility, rooting its decentralization in crypto-native values \citep{silberling_farcastercryptobased_2024,weinstein_controlcommunity_2024}. Instead of \Ser\ owners commissioning communication between one another, or a standalone actor setting up an \Agg, Farcaster \Ser s are designed to automatically connect and synchronize data between every other \Ser\ on the network. To do this, Farcaster uses Gossipsub, a distributed data-sharing system. Gossipsub is a separate message propagation protocol that Farcaster leverages to ensure that data is quickly disseminated to all \Ser s in the network , even as it grows in size \cite{farcaster_architecturefarcaster_2024}. Each \Ser\ thus stores the full contents of the global \Spa, independently compiling and directly relaying data without intermediaries. This makes running a \Ser\ on Farcaster noticeably expensive \cite{farcaster_scalingfarcaster_2024}. Meanwhile, multiple \Cli s can interact with a given \Ser. This results in much of the curatorial influence with Farcaster existing within \Cli s, with many implementations also taking a hands off approach with \SCu s, opting for composable moderation. To join, users generate accounts tied directly to their Ethereum-based crypto wallet, and metadata about those accounts is anchored to the blockchain \cite{farcaster_scalingfarcaster_2024}. This not only ensures decentralized control of identity and access (through private wallet keys), but also enables \ISt\ interoperability across applications outside the Farcaster ecosystem---including Ethereum-based services like {Polymarket}\footnote{\texttt{https://polymarket.com/}} and {Opensea}\footnote{\texttt{https://opensea.io/}} \cite{gabriele_futurefarcaster_2024}. The fourth column of Tables \ref{tab:architectural_components_comparison} and \ref{tab:interaction_components_comparison} summarizes the architectural and interaction components of this protocol.

\subsection{Power and Trade-offs in Protocol Design}
In what follows, we surface three key issues of power as we compare the protocols' distinct approaches to decentralizing social media through the lens of our framework. Drawing on insights from interviews, we contextualize the protocols’ design choices and identify distinct challenges that arise in how power is allocated.

\subsubsection{Power Over Identity}
A core question for any decentralized social media protocol is how people are able to manage their identity on it: creating, migrating, deleting an account, as well as the information associated with it. 

AT Protocol and ActivityPub allow users to choose a handle, set a password, and become a member of a \Spa\ within minutes via a \Cli\ with a familiar graphical user interface. AT Protocol aims to replicate the experience of centralized social media, which has helped make platforms like Bluesky approachable for the average user \cite{lemmer-webber_activitypubw3c_2018}. While ActivityPub has been noted to be a bit more challenging \citep{silberling_beginnersguide_2023}---primarily due to confusion around choosing a \Ser/\Spa\ to become a member of---it has nonetheless grown into one of the most widely used systems for decentralized social media \citep{he_flockingmastodon_2023, jeong_exploringplatform_2024,lunden_howmastodon_2022,sherr_twitterwho_2023} thanks to its familiar, centralized-style way of managing user accounts at the \Ser\ level. In contrast, with Nostr and Farcaster, users generate and manage their own identity keys. From the moment a new Nostr user creates an account through any available \Cli, they are assigned an alphanumeric string as their identity key, which they use to access their \ISt. Farcaster binds identity to data stored in what is essentially a distributed database (the Ethereum blockchain), requiring users to pay roughly \$5 from their corresponding cryptocurrency wallet to register a new identity. While these approaches increase user agency by allowing users to truly own their \ISt\ independently, they increase the cost and knowledge barriers for who can easily join a given \Spa.

\textbf{When ownership and responsibility for \ISt\ is separated from the \Ser, everyday users have more power over identity---but must have sufficient technical knowledge to do so, and well}. This is exemplified when we further consider how personal data is subsequently managed after sign-up across the protocols. In ActivityPub and AT Protocol, the identity keys associated with \ISt\ typically reside within the \SSt\ of a \Ser, which means that the \Ser\ owners---whoever that may be---hold the keys to authentication and recovery \cite{robertson_twittersopensource_2022}. This makes account access simple, but also renders users vulnerable to \Ser\ failure, policy shifts, or unilateral bans. Recent cases \citep[see][]{mir_fbiseizure_2023} and research highlights how \Ser\ maintenance, especially around privacy and security \citep{tosch_privacypolicies_2024,hwang_trustfriction_2025}, can be consequential and ends up relying on a small subset of actors functioning as \Ser\ admins. AT Protocol's idea of ``credible exit'' (where users can manually export their identity and keys from one \Ser\ to another) aims to address this, but now requires active management and technical knowledge \cite{atprotocol_atprotoethos_2025}. Nostr and Farcaster separate identity even more, making \ISt\ independent. Users are not dependent on any single \Ser to validate their presence in a \Spa, and they can switch \Ser s without losing their social identity. However, this also means that losing identity keys is irreversible, in that they cannot be retrieved by asking a \Ser\ admin. 
P2 noted that the relative ease of account access and identity management on AT Protocol, as most usually rely on the \Ser\ owner, made it much more accessible to people, compared to other protocols: ``\textit{(It's) a huge driver for people to come and adopt it ... (and) convince others to adopt it ... (and why) ... Bluesky (and by extension AT Protocol) is winning.}'' 

In sum, as identity is decoupled from \Ser s and into the hands of users, so too does the responsibility for maintaining it. Protocols like ActivityPub and AT Protocol centralize identity management by storing user identity keys on \Ser s, relieving users of the burden of key maintenance, but also granting \Ser\ operators substantial influence over the broader \Spa s. In contrast, protocols like Nostr and Farcaster offer greater user autonomy by allowing individuals to directly control their identity keys. However, this raises risks as individuals must have the resources and knowledge to manage these identity keys well.

\subsubsection{Power Over Curation}
As flows of user-generated \textsc{Actions} become \textsc{Feeds}, each protocol also contends with power over curation: who can rank and organize information on decentralized social media, and how. Across our four protocols, \Cli s often determine the algorithm for the \Fee. As a result, the \Cli\ has the potential to play a substantial role in curation. With Farcaster, all \Ser s share data in the same standardized way and are effectively interchangeable, and curatorial power is concentrated almost entirely at the \Cli\ level. \Cli s are typically developed and maintained by teams who tend to take a relatively hands-off approach to moderation\cite{a16z_a16zawesomefarcaster_2024, gabriele_futurefarcaster_2024}.
While AT Protocol similarly places most curatorial influence at the \Cli\ layer, its architecture gives \Ser s a slightly more active role. As mentioned earlier, \Agg s work to equalize discrepancies in data access and visibility by relaying content across all \Ser s in the network. However, \Ser s can influence the visibility of \Act s by blocking or filtering data sent to or received from \Agg s \cite{atproto_selfhosting_2025, bnewbold.net_notesrunning_2024}, and individual users have significantly enhanced \PCu\ mechanisms \citep{bluesky_blueskysstackable_2024}. However, \Cli s remain the most consequential actors in this respect. With one main \Cli\ being used in practice with AT Protocol, P9 (one of the engineers at Bluesky PBC) acknowledged concerns about centralization: ``\textit{If we went shitty (we'd hope) someone else could start another one}.''
For both Farcaster and AT Protocol, the concentration of power over curation around \Cli s can result in \textit{de facto} re-centralization, along with a more consistent experience of content curation---and moderation---as a result.

We note that AT Protocol also gives curatorial powers to other components. \textbf{Power over curation can be decentralized by distributing the ability to curate content across multiple components, not just \Cli s---especially if components are owned independently. However, this can raise new challenges around consistency and coordination.} To elaborate, we can consider the cases of ActivityPub and Nostr, which distribute power over curation by giving moderation abilities to both \Cli s and \Ser s. In ActivityPub, independent \Ser\ administrators act as moderators, with significant discretion over local content policies and inter-server relationships \cite{nicholson_mastodonrules_2023,colglazier_effectsgroup_2024}. Nostr takes this even further by leaving content decisions entirely to independent \Ser\ \textit{and} \Cli\ operators---reflecting its minimal moderation stance in favor of user expression, which has led one of its \Cli s to being banned in China \citep{liao_damuspulled_2023}. 
This has led to inconsistent moderation standards across implementations of both protocols. Recent work on ActivityPub instances notes how inconsistent practices can result in new risks for harm, such as privacy violations \citep{hwang_trustfriction_2025}. P3 compared Nostr to \textit{Burning Man}, a countercultural festival rooted in radical self-expression, saying ``\textit{art was just created by groups of people, and you could just show up; \dots it wasn't always directed \dots (but) that’s what Nostr is like}.'' 

Ultimately, Farcaster and AT Protocol see concerns about concentration of power over curation because of the dominance of one component, the \Cli, although the latter begins to enable curation through other components as well. Meanwhile, while ActivityPub and Nostr offer a more decentralized and pluralistic approach by distributing power over curation to independently run \Cli s and \Ser s, they also introduce challenges around coordination and consistency. Because \Spa\ curation is shared between \Cli\ and \Ser\ operators in these protocols, the experience can feel uneven: less prone to centralized control, but also more fragmented, possibly increasing the risk of exposing users to content they find harmful or unwelcome \citep{hwang_trustfriction_2025,tosch_privacypolicies_2024}.

\subsubsection{Power Over Infrastructure}
Finally, the dynamics of identity and curation are deeply influenced by power over infrastructure---that is, who has the ability implement the protocol to create \Spa s that can interoperate with the broader social ecosystem.

For protocols like ActivityPub and Nostr, infrastructure tends to be more accessible in terms of resources: individuals can spin up a new \Spa\ by operating a \Cli\ and \Ser. In ActivityPub's case, they can do so using pre-packaged, standardized software like Mastodon or Pixelfed. This ease of entry lowers barriers to participation and encourages a broad base of infrastructure operators. However, because \Cli s, \Ser s, and \Spa s are often bundled together, power can become concentrated in the hands of admins who now have purview over all three architectural components. P3 noted that ActivityPub \Ser\ owners ``\textit{...have a lot more power and control with ActivityPub (compared to other protocols)},'' while P7 described the protocol as ``\textit{semi-decentralized \dots with points where tons of information (and power) is stored}.''
While Nostr takes a similar architectural approach, it requires more technical knowledge to set up infrastructure like \Cli s and \Ser s, which do not often come bundled like with ActivityPub \cite{neves_howsetup_2022, pluja_settingnostr_2025}. This openness offers greater flexibility, allowing \Cli s and \Ser s to mix and match but requiring more technical knowledge about which to choose. Moreover, operating any \Spa\ demands time and resources \citep{anaobi_willadmins_2023}. Because both protocols still require some amount of technical background, most users join existing \Spa s. Faced with a huge number of possible \Spa s to join, users often gravitate toward the most visible or active \Spa s, which gain traction because of network effects. P4 noted: ``\textit{you need to be connected to the correct [\Ser] to see stuff (you wish to see)},'' stressing how discoverability and visibility shape infrastructural dominance even in flat, decentralized designs. This challenge is not unique to Nostr---ActivityPub ecosystems face similar dynamics, with large \Spa s like \texttt{mastodon.social} emerging as central gathering points.

\textbf{Power over infrastructure can be decentralized by making it possible for users to create and own architectural components. However, as we observed with ActivityPub and Nostr, the resulting systems still face new risks of re-centralization in practice due to constraints in knowledge, resources, and time. Further, they may come with considerable financial costs for people.} AT Protocol and Farcaster highlight how resource costs of running infrastructure can prevent users from operating their own architectural components as envisioned by the protocols. 
To fully own an AT Protocol \Spa, one must not only run a \Ser\ and \Cli, but also operate an independent \Agg---a role typically filled by Bluesky PBC. Running an \Agg\ carries significant technical and financial burdens due to the heavy compute required to index and serve the network’s data. Without this, developers remain dependent on Bluesky PBC's infrastructure \cite{newbold_replybluesky_2024}. Farcaster's arrangement of architectural components initially suggests a return to a simpler workflow, requiring only a \Cli\ and \Ser\ \cite{weinstein_controlcommunity_2024}. However, each \Ser\ must also handle data replication and indexing, making it just as computationally demanding as running an \Agg.
The high technical and financial barriers required to meaningfully participate in infrastructure ownership create a new kind of centralization: only well-resourced actors can take on these roles. 
As a result, despite the decentralized architectures stipulated by the protocols, companies like Bluesky PBC and the team behind Warpcast still wield outsized influence in practice---an outcome that ironically mirrors the kinds of centralizing dynamics these systems were designed to avoid.

In short, decentralizing power over infrastructure depends heavily on the barriers to entry associated with architectural components ownership. ActivityPub and Nostr make it relatively easy for individuals to launch new \Spa s, especially with software like Mastodon that bundles \Cli s and \Ser s. However, resource and knowledge constraints lead most users to join existing \Spa s, and the two protocols see dominance by a few highly-visible instances---such as \texttt{mastodon.social} or Damus---that become \textit{de facto} central points in the network. Meanwhile, running a Farcaster \Ser\ or AT Protocol \Agg\ requires substantial technical and financial investment, preventing decentralization of power over infrastructure in practice.

\section{Discussion} \label{s_discussion}
Although protocols are sets of rules, they become tangible artifacts when they are implemented. We presented a framework that can be used to describe how a protocol is operationalized into a distinct arrangement of socio-technical components, with the aim of providing a consistent lexicon by which users, designers, and developers can ``see'' protocols. Our application of the framework to describe four well-known protocols---ActivityPub, AT Protocol, Nostr, Farcaster---underscored how these protocols take different approaches to ``decentralization'' and, in turn, make distinct trade-offs through protocol design. In particular, we consider the politics of protocols: the ways that protocols assign and enable ownership of components by different actors. 

Looking across issues of power over identity, curation, and infrastructure, we note that decentralized social media protocols can take multiple approaches to decentralization: 
\begin{itemize}
    \item \textbf{Separation of ownership of certain components} can decentralize power over identity;
    \item \textbf{Distributing the ability to do certain tasks, e.g., moderate and filter, across components} can decentralized power over curation, especially if those components are operated by different actors; 
    \item \textbf{Making it easier for people to create and operate their own instantiations of components} (particularly architectural components) can decentralize power over infrastructure.
\end{itemize}

\noindent Attempts to decentralize around each issue tried to shift more power to users in some way. However, for each issue, we also observe that the approach highlighted faced substantial challenges. With power over identity, users need to have the requisite technical knowledge to manage their identity keys for their \ISt. With power over curation, having multiple, independent points of decision-making of content raises the potential for bad actors in the network to propagate harmful content. With power over infrastructure, technical, resource, time, and cost barriers can be serious roadblocks to independent actors operating architectural components, resulting in multiple potential paths to \textit{de facto} centralization. 

Across the four protocols we examined, each had a unique take on how it combined and prioritized the three approaches to decentralization listed above. Each protocol thus also faces distinct challenges in how it aims to operationalize decentralization, making trade-offs between decentralization and the associated risks, barriers, and/or costs for each approach. Prior work on decentralized social media has often focused on user behavior within individual implementations of protocols---particularly \texttt{mastodon.social}---and frequently treat decentralization as a given \citep{cava_driverssocial_2023, bono_explorationdecentralized_2024}. While valuable in articulating current experiences of users on decentralized social media platforms (e.g., challenges in decentralization), our works shows how protocols make key decisions about \textit{how} to decentralize that are likely to give rise to those dynamics documented in recent empirical work, such as challenges faced by admins in the ActivityPub network \citep{anaobi_willadmins_2023,zhang_troubleparadise_2024} as they take on key roles as operators of \Spa s, \Ser s, \textit{and} \Cli s. More broadly, as each protocol makes different trade-offs, we underscore that what decentralization looks like---and should look like---in social media is still under contention and can be deeply consequential. 

Accordingly, we make a call for social computing researchers to attend to protocol design more closely, extending the current focus on platform experiences in the growing domain of decentralized social media research. 
Focusing on protocols reveals not only how decentralization is enacted, but also which values it operationalizes, for whom, and with what consequences. As Winner famously observed, artifacts have politics \citep{winner_artifactshave_1980}; in decentralized social media, protocols are the artifacts that shape who holds power over identity, data, moderation, and access. Through our conceptual framework, we aim to help surface these politics by breaking protocols down into a set of components that can help researchers see the socio-technical artifact a protocol defines, which can then be evaluated, compared, and contextualized within its normative goals. 

For example, while our observations about concerns relating to power over identity and curation largely reflect the design choices of the protocols, concerns about power over infrastructure highlight a gap between what the protocols envision and what happens in practice. As such, a key area for future work is both understanding why such gaps emerge and how they might be mitigated---whether through better tooling, more sustainable funding models, or clearer pathways for non-experts to operate core components.
Our analyses of power over identity, curation, and infrastructure offer some insight into how these approaches to decentralization play out. They also do not indicate that they are mutually exclusive; future work might explore how different approaches to decentralization may be deliberately combined to counterbalance the limitations of any single approach. For example, separating identity from infrastructure while distributing curation tasks may offer users more control without requiring full technical self-sufficiency. In this vein, we argue that protocol design can be a site of socio-technical imagination: one that articulates not only how systems work, but also who they empower and how. By foregrounding protocol design as an analytical and design space, we hope to encourage researchers, designers, and builders to more deliberately engage with the values encoded into the infrastructures of decentralized social media---and to interrogate what kinds of decentralization we ultimately want to build.

\section{Conclusion}
Decentralized social media protocols have re-emerged in recent years as a potential solution to the centralized platforms that have dominated the past two decades. However, protocols differ substantially, reflecting distinct values and making trade-offs in their approaches to decentralization. In our work, we presented a novel conceptual framework to help ``see'' the politics of protocols, providing a lexicon of components that make up the social media system envisioned by the protocol. Describing four major real-world protocols, we leveraged this framework to surface how protocols allocate decision-making power across components. Our analysis shows that decentralization is not a fixed end state but a series of consequential design decisions that distribute control in varied and sometimes contradictory ways. Rather than revealing a singular ``best'' model, turning our attention to protocol design raises questions of what kinds of decentralization we want, for whom, and under what conditions.

To move this conversation forward, researchers and developers must turn their attention to protocol design. We argue that protocol design is not just a technical question but a political one about who gets to set the rules of online life. Moving beyond binary framings of ``centralized vs. decentralized,'' we call for more grounded, comparative, and values-aware approaches to studying and building these systems. Future work should continue mapping and evaluating how different protocol components interact, tracing unintended consequences, and surfacing where ideological commitments diverge from lived experience. Drawing on these insights, designers can identify weaknesses in decentralized systems and devise tools that may improve user autonomy, working towards improved protocols that help us better realize the promise of decentralization for a healthier digital world.

\newpage
\bibliographystyle{ACM-Reference-Format}

\newpage
\appendix

\end{document}